\documentclass[final,5p,times,twocolumn]{elsarticle}

\usepackage{amssymb}
\usepackage{hyperref}
\usepackage[dvipsnames]{xcolor}
\usepackage{xspace}
\usepackage{siunitx}
\usepackage{physics}
\usepackage{multirow}
\usepackage{mhchem}

\newcommand*{\red}[1]{{\color{red} #1}}

\newcommand*{\duo}{\textsc{Duo}\xspace}
\newcommand*{\eg}{\textit{e.g.}\xspace}
\newcommand*{\etal}{\textit{et al.}\xspace}

\newcommand*{\ie}{\textit{i.e.}\xspace}

\newcommand*{\abinitio}{\textit{ab~initio}\xspace}
\newcommand{\exocross}{\textsc{ExoCross}\xspace}
\newcommand*{\Ldoubling}{{$\varLambda$\,-\,doubling}\xspace}
\newcommand*{\cm}{cm$^{-1}$}

\newcommand*{\VOXSigma}{\mathrm{X}\,^4\Sigma^-}
\newcommand*{\VOApPhi}{\mathrm{A}'\,^4\Phi}
\newcommand*{\VOAPi}{\mathrm{A}\,^4\Pi}
\newcommand*{\VOBPi}{\mathrm{B}\,^4\Pi}
\newcommand*{\VOCSigma}{\mathrm{C}\,^4\Sigma^-}
\newcommand*{\VODDelta}{\mathrm{D}\,^4\Delta}

\newcommand*{\VOdouDelta}{\mathrm{1}\,^2\Delta}
\newcommand*{\VOdouSigmap}{\mathrm{1}\,^2\Sigma^+}
\newcommand*{\VOdouPhi}{\mathrm{1}\,^2\Phi}
\newcommand*{\VOdouFirstPi}{\mathrm{1}\,^2\Pi}
\newcommand*{\VOdouSecondPi}{\mathrm{2}\,^2\Pi}

\journal{J. Molec. Spectrosc.}

\begin{document}

\begin{frontmatter}

\title{An empirical spectroscopic model for eleven electronic states of VO}

\author[inst1]{Qianwei Qu, Sergei N. Yurchenko and Jonathan Tennyson}

\affiliation[UCL]{organization={Department of Physics and Astronomy, University College London},
            city={London},
            postcode={WC1E 6BT}, 
            country={United Kingdom}}

\begin{abstract}
Previously-determined empirical energy levels are used to construct a rovibronic model for the 
$\VOXSigma$, $\VOApPhi$, $\VOAPi$, $\VOBPi$, $\VOCSigma$, $\VODDelta$, $\VOdouDelta$, $\VOdouSigmap$, $\VOdouPhi$, $\VOdouFirstPi$   and $\VOdouSecondPi$   electronic states of vanadium mononoxide. The spectrum of VO is characterized by many couplings and crossings between
the states
associated with these curves. The model is based on the use of potential energy curves,
represented as extended Morse oscillators,
and couplings (spin-orbit, spin-spin, angular momentum), represented by polynomials, which are tuned to the data plus an empirical allowance for spin-rotation couplings. The curves
are used as input for the variational nuclear motion code \duo. For the
$\VOXSigma$, $\VOdouPhi$ and  $\VOdouFirstPi$ states the model reproduces the observed energy to significantly better than 0.1 \cm. For
the other states the standard deviations are between 0.25 and 1.5 \cm. Further experimental data and consideration of hyperfine effects would probably be needed to significantly
improve this situation.
\end{abstract}

\begin{keyword}
Variational calculations \sep potential energy curves \sep spin-orbit coupling
\PACS 0000 \sep 1111
\MSC 0000 \sep 1111

\red{The manuscript has been accepted by the \emph{Journal of Molecular Spectroscopy}}
\end{keyword}

\end{frontmatter}


\section{Introduction}
\label{sec:sample1}
Vanadium monoxide (VO) is a diatomic molecule with
a complicated open shell structure \cite{89Merer.VO,jt873}; it absorbs strongly at red and
near infrared wavelengths. VO is thought to provide an important source
of opacity in the atmospheres of both cool stars  \cite{09Bernath.VO} and  of exoplanets \cite{08FoLoMa.VO,10MaSexx.VO}. While VO bands are well-know
in the spectrum of cool stars \cite{11TyKaSc.VO} and sunspots \cite{08SrBaRa.VO}, only 
tentative detections have been made in the atmospheres of exoplanets 
\cite{16EvSiWa.VO,17TuLeBi.VO,17PaChPr.VO,jt699,20GoMaDrSi.VO,20LeWaMa.VO,21ChEdxx.exo}.
The charged species \ce{VO^{-}} and \ce{VO^{+}} are also important targets 
of theoretical and experimental studies,
see \eg, Miliordos and Mavridis \cite{07MiMaxx.VO} and 
the references therein.

Theoretically modelling of VO spectra is a challenge both from the perspective
of electronic structure calculations \cite{jt783,07MiMaxx.VO,jt623,21JiChBo.VO} and,
as will be seen below, from the perspective of representing the nuclear
motion due to many resonances between different states.
McKemmish \etal \cite{jt644} constructed the so-called VOMYT line list for VO as part
of the ExoMol project \cite{jt528,jt810}. 
The VOMYT line list probably provides a reasonable
basis for modelling VO opacities 
but has been demonstrated to be inadequate for
high resolution studies of exoplanetary spectra \cite{22DeKeSn.VO}.

The spectrum of VO has been well-studied experimentally 
\cite{57LaSe,81HoMeMi,82ChHaMe,82ChTaMe,87MeHuChTa,91SuFrLoGi,
94ChHaHuHu,95AdBaBeBo,97KaLiLuSa,02RaBeDaMe,05RaBe,08FlZi,09HoHaMa}, 
see the summary by Bowesman \etal \cite{jt869};
however, the complexity of its spectrum
means that there are still important gaps in these studies. 
Although, as discussed below,
our general approach to constructing a spectroscopic model for open shell molecules
such as VO is based on variational solutions of the nuclear motion Schrodinger equation,
our work on VO \cite{jt873,jt869} and other open shell molecules 
\cite{jt760,jt835,jt847,jt875} has made extensive use of Western's effective
Hamiltonian program PGOPHER \cite{PGOPHER}.  We favour the method based on variational
solutions as it usually shows much more stable extrapolation properties which is an 
important feature when one wishes to compute line lists for hot molecules.

\begin{figure}
    \centering 
    \includegraphics{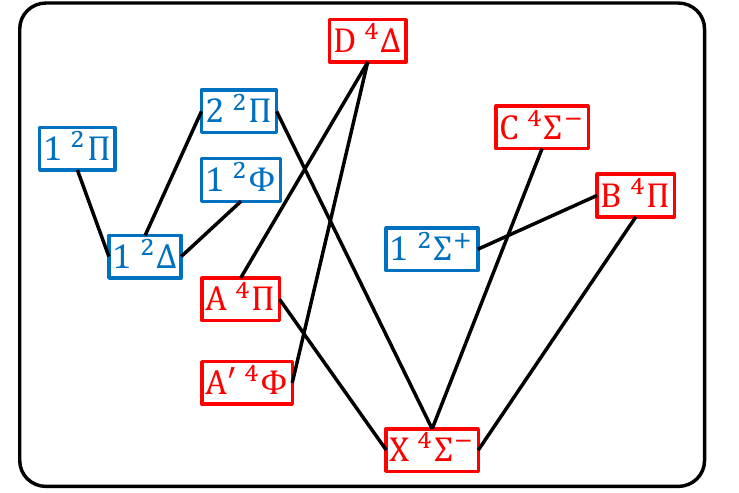}
    \caption{Band systems connecting the
     11 electronics included in the 
     model of this work.
     See Bowesman \etal \cite{jt869}
     for a more comprehensive diagram.}
    \label{fig:bandsystems}
\end{figure}

In this paper, we attempt to build a spectroscopic model
which covers the most important low-lying states of VO, 
including 6 quartet and 5 doublet electronic states; given the difficulty in fully characterizing the
couplings between these state we limit ourselves to
considering only the coupling terms necessary to provide a reasonable spectroscopic model.
The 11 electronic states considered are shown
in Fig.\,\ref{fig:bandsystems}; these state are all
part of the VOMYT model.
They are responsible for the main
absorption bands in the near infrared and visible, 
or are states which are known to
interact with them.

The purpose of this work 
is to provide the basis for a more reliable model than VOMYT
as well providing 
for the quantum number assignments for unassigned 
recent observations \cite{09HoHaMa,jt869}.
Thus, 
our focus  is on the accuracy of energy levels
and we develop an empirical model that excludes 
most of the off-diagonal couplings 
considered in VOMYT \cite{jt644}.
In the recent study by Bowesman \etal \cite{jt869},
a spectroscopic network of VO transitions was constructed
as a prelude to a MARVEL 
(measured active rotation vibration energy levels) \cite{jt412} study of 
the 11 states involved in this network.
The next two sections will demonstrate
our effort to reproduce the rovibronic energy levels
obtained from the MARVEL study.
The other two states ($\mathrm{a}\,^2\Sigma^-$ and
$\mathrm{b}\,^2\Gamma$ ) 
included in VOMYT \cite{jt644} were not analyzed in
the work of Bowesman \etal \cite{jt869}
and are, therefore, excluded from our model as well.

\section{Spectroscopic model}

The model was developed using the open-source program 
for variational calculation of diatomic spectra, \duo\footnote{\href{https://github.com/exomol/Duo}{https://github.com/exomol/Duo}} \cite{jt609}.
All nuclear motion calculations used 10 vibrationally contracted basis functions
for each electronic states based on 401 sinc-DVR grid points,
covering the internuclear distance range
from 1.2 to 4 \si{\angstrom}. 
The upper bound of the energy 
was set to \SI{50000}{\per\cm},
however energy levels of interest for this work 
are expected to be below \SI{21000}{\per\cm}.
The number of vibrational levels considered for 
each electronic state
is listed in Table\,\ref*{tab:vibnumber}.

\begin{table}[htbp]
    \centering
    \caption{Vibrational states considered  for 
    each electronic state.
    Table\,5 of Bowesman \etal \cite{jt869}
    lists the corresponding experimental studies.}
    \label{tab:vibnumber}
      \begin{tabular}{cll}
        \hline
      Number   & States & $v$ \\
      \hline
      1    & $\VOXSigma$         & 0-2 \\
      2    & $\VOApPhi$          & 0-2 \\
      3    & $\VOAPi$            & 0 \\
      4    & $\VOBPi$            & 0-1 \\
      5    & $\VOCSigma$         & 0-2 \\
      6    & $\VODDelta$         & 0-1 \\
      9    & $\VOdouDelta$       & 0-1 \\
      10   & $\VOdouSigmap$      & 2-3 \\
      11   & $\VOdouPhi$         & 0 \\
      12   & $\VOdouFirstPi$     & 0-3 \\
      13   & $\VOdouSecondPi$    & 0-1 \\
      \hline
      \end{tabular}
  \end{table}%

\subsection{Potential energy curves}
The potential energy curves (PECs)
of all states are represented by second-order
extended Morse oscillator (EMO) function \cite{EMO}
\begin{equation}
    V(R)=T_{\mathrm{e}}+\left(D_{\mathrm{e}}-T_{\mathrm{e}}\right)
    \left[1-\exp \left(
        -\beta_{\mathrm{EMO}}(R)
        \left(R-R_{\mathrm{e}}\right)
    \right)
    \right]^{2}, 
    \label{eq:emo}
\end{equation}
where $R$ is the internuclear distance, $R_{\mathrm{e}}$ is the equilibrium bond length and
$T_{\mathrm{e}}$ is the potential minimum;
$\beta_\mathrm{EMO}$ is given by the formula:
\begin{equation}
    \beta_{\mathrm{EMO}}(R)= b_0 + b_1\, y_p(R)
    +b_2\, y_p^{2}(R),
\end{equation}
where $y_{p}(R)$ is:
\begin{equation}
    y_{p}(R)=\frac{R^{p}-(R_\mathrm{e})^{p}}
    {R^{p}+(R_\mathrm{e})^{p}} \,.
    \label{eq:surkus}
\end{equation}

\duo's users can set independent integer values of 
$p$ on the left- and right-hand
sides of $R_{\mathrm{e}}$
and we chose $p=4$ for both sides.
The parameters of fitted EMO functions 
are listed in Table\,\ref{tab:emoparameters}.

\begin{table*}[h]
\centering
\caption{Optimized extended Morse oscillator (EMO) parameters 
for the potential energy curves, see Eq.\eqref{eq:emo}.}
\label{tab:emoparameters}
\begin{tabular}{clrrrrr}
    \hline
    Number & State & $T_{\mathrm{e}}$ [\si{\per\cm}]  & $R_{\mathrm{e}}$ [\si{\angstrom}]   & $b_0$ [\si{\per\angstrom}]   & $b_1$  [\si{\per\angstrom}]   & $b_2$ [\si{\per\angstrom}]  \\
    \hline
    1  & $\VOXSigma$        & 0 & 1.58947260 & 1.87183393 & \num{6.04260663E-03} & \num{-1.15873102E-01} \\
    2  & $\VOApPhi$         & 7292.4851 & 1.62591915 & 1.88461684 & \num{-1.33209148E-02} & \num{1.58520316E-01} \\
    3  & $\VOAPi$           & 9534.8342    & 1.63405670 & 1.92275217 &       &  \\
    4  & $\VOBPi$           & 12659.4107 & 1.64032109 & 1.93810207 &    0   & \num{-2.27321332E-01} \\
    5  & $\VOCSigma$        & 17494.4977 & 1.67098029 & 1.94981612 & \num{-5.17057888E-02} & \num{3.27120699E-01} \\
    6  & $\VODDelta$        & 19238.8720 & 1.68337218 & 1.91485252 &     0  & 3.03001473 \\
    7  & $\VOdouDelta$      & 9867.6920 & 1.58225344 & 2.06276090 &   0    & \num{-4.15515802E-01} \\
    8  & $\VOdouSigmap$     & 10376.3396 & 1.57856000 & 2.15102878 &    0   & \num{-3.66304253E-01} \\
    9  & $\VOdouPhi$        & 15442.9452 & 1.62904530 & 2.09925561 &       &  \\
    10 & $\VOdouFirstPi$    & 17099.9353 & 1.62955854 & 2.10656122 & \num{2.73289447E-01} & \num{1.64122298E-01} \\
    11 & $\VOdouSecondPi$   & 18139.4421 & 1.62244306 & 2.15748234 &       &  \\
    \hline
\end{tabular}
\end{table*}
  
\subsection{Diagonal couplings within a electronic state}

All coupling curves are represented by low-order polynomials
\begin{equation}
    p(R) = a_0 + \sum_{n=1}^{\dots} a_n\, (R-R_\mathrm{e})^n.
    \label{eq:poly}
\end{equation}
Our aim is to use the fewest terms to satisfactorily reproduce 
the experimental energy levels.
Thus,
we started with uncoupled PECs
for each state and then,
added those couplings curves necessary to
reduce the fitting errors.

We assume the spin splittings of $\VOXSigma$ and $\VOCSigma$ mainly depend on spin-spin coupling terms.
Although the assumption is not supported by \abinitio calculations, 
\eg see our previous discussion of the $\VOXSigma$ state\cite{jt873},
this empirical representation can usually give accurate energy levels.
Apart from the spin-spin interaction,
spin-rotation terms were used to resolve $J$-dependent
spin splittings of $^2\Sigma$ and $^4\Sigma$ states,
where $J$ is the total angular momentum.

In contrast,
spin-orbit terms dominate the spin splittings of 
$\Pi$, $\Delta$, $\Phi$ and $\Gamma$ states
and are $J$-dependent too.
Therefore,
we replaced spin-rotation curves with
spin-orbit coupling ones for those states.
Spin-spin coupling curves were still used for 
the quartet states, 
\ie $\VOApPhi$, $\VOAPi$, $\VOBPi$ and $\VODDelta$.

The optimized parameters of
diagonal spin-spin, spin-rotation and spin-orbit couplings
are listed in Tables \ref{tab:spinspin}, \ref{tab:spinrotation}
and \ref{tab:diagonalso}, respectively.

\begin{table*}[h]
\centering
\caption{Optimized polynomial parameters for diagonal spin-spin coupling curves.}
\label{tab:spinspin}
    \begin{tabular}{lrrr}
    \hline
    State & $a_0$ [\si{\per\cm}]     & $a_1$  [\si{\per\cm\per\angstrom}]  & $a_2$ [\si{\cm^{-1}\angstrom^{-2}}]\\
    \hline
    $\VOXSigma$     & 2.05042406     & 2.76918955    &  \\
    $\VOApPhi$    & -1.08746355        &       &  \\
    $\VOAPi$     & 1.89104909           &       &  \\
    $\VOBPi$     & 2.62698261        &       &  \\
    $\VOCSigma$     & 0.73293575        &       &  \\
    $\VODDelta$     & 1.29783687    &     0  & \num{-6.22043554E+02} \\
    \hline
    \end{tabular}
\end{table*}

\begin{table}[h]
\centering
\caption{Optimized polynomial parameters for diagonal spin-rotation coupling curves.}
\label{tab:spinrotation}
    \begin{tabular}{lrr}
    \hline
    State & $a_0$ [\si{\per\cm}]     & $a_1$ [\si{\cm^{-1}\angstrom^{-1}}] \\
    \hline
    $\VOXSigma$    & \num{2.14387841E-02}   & \num{1.58712788E-01} \\
    $\VOCSigma$     & \num{-1.83464308E-02}       &  \\
    $\VOdouSigmap$  & \num{-7.90000000E-03}      &  \\
    \hline
    \end{tabular}
\end{table}

\begin{table*}[h]
    \centering
    \caption{Optimized polynomial parameters for diagonal spin-orbit coupling curves.}
    \label{tab:diagonalso}
    \begin{tabular}{lrrrrr}
    \hline
    State & $a_0$ [\si{\per\cm}]     & $a_1$  [\si{\per\cm\per\angstrom}]  & $a_2$ [\si{\cm^{-1}\angstrom^{-2}}]   & $a_3$[\si{\cm^{-1}\angstrom^{-3}}] \\
    \hline
    $\VOApPhi$   & \num{2.60752743E+02}  & \num{-3.08377540E+02} & \num{-3.01987104E+03} & \num{5.20656186E+04} \\
    $\VOAPi$     & \num{5.26495455E+01}           &       &       &  \\
    $\VOBPi$     & \num{9.70343261E+01}  & \num{-3.31204551E+01} &       &  \\
    $\VODDelta$     & \num{1.39054588E+02}  &    0   & \num{2.59920512E+03} &  \\
    $\VOdouDelta$ & \num{1.80055832E+02}  & \num{-1.58772300E+02} &       &  \\
    $\VOdouPhi$   & \num{1.86052851E+02}        &       &       &  \\
    $\VOdouFirstPi$    & \num{1.24587433E+02} & \num{8.26877879E+01} & \num{-2.70479601E+02} &  \\
    $\VOdouSecondPi$    & \num{6.03714458E+01}        &       &       &  \\
    \hline
    \end{tabular}
\end{table*}

We also modeled the \Ldoubling of $\Pi$ states
by using the empirical terms defined
by Brown and Merer \cite{79BrMexx.methods}:
\begin{align}
    &\mel**{\mp 1, \varSigma \pm 2, J, \varOmega}
        {\hat{H}_\mathrm{LD}}
        {\pm 1, \varSigma, J, \varOmega}
        =  \frac{1}{2}\left(o_{v}+p_{v}+q_{v}\right) \times \notag \\
        &    \sqrt{[S(S+1)-\varSigma(\varSigma \pm 1)]
            [S(S+1)-(\varSigma \pm 1)(\varSigma \pm 2)]} \, ,
        \label{eq:Lopq}\\
    &\mel**{\mp 1, \varSigma \pm 1, J, \varOmega \mp 1}
        {\hat{H}_\mathrm{LD}}{\pm 1, \varSigma, J, \varOmega}
        =  -\frac{1}{2}\left(p_{v}+2 q_{v}\right) \times \notag \\
    &   \sqrt{[S(S+1)-\varSigma(\varSigma \pm 1)]
            [J(J+1)-\varOmega(\varOmega \mp 1)]} \, ,
        \label{eq:Lp2q}\\
    & \mel**{\mp 1, \varSigma, J, \varOmega \mp 2}{\hat{H}_\mathrm{LD}}
        {\pm 1, \varSigma, J ,\varOmega}
        = \frac{1}{2} q_{v} \times \notag \\
    &   \sqrt{[J(J+1)-\varOmega(\varOmega \mp 1)]
            [J(J+1)-(\varOmega \mp 1)(\varOmega \mp 2)]} \, .
        \label{eq:Lq}
\end{align}
In these equations,
$o$, $p$ and $q$ depend on vibrational quantum number, $v$,
as Brown and Merer derived the equations based on effective Hamiltonians.
In our variational model,
we use coupling curves instead.
For quartet states,
we include all the three coupling curves,
while for doublet states,
Eq.\,\eqref{eq:Lopq} is automatically zero
and we only need the curves of $p+2q$ and $q$.
The fitted parameters for the $o+p+q$, $p+2q$ and $q$ curves
are listed in Table \ref{tab:Lambda}.

\begin{table}[h]
\centering
\caption{Optimized polynomial parameters for the $o+p+q$, $p+2q$ and $q$ curves due to \Ldoubling coupling.}
\label{tab:Lambda}
    \begin{tabular}{clr}
    \hline
    Coupling & State & $a_0$ [\si{\per\cm}]\\
    \hline
    \multirow{2}[0]{*}{$o+p+q$ } &    $\VOAPi$      & 2.05253781    \\
    &$\VOBPi$ & 1.20113529   \\
    \hline
    \multirow{4}[0]{*}{$p+2q$ } &    $\VOAPi$      & \num{-1.41692302E-02}    \\
    & $\VOBPi$     & \num{3.18974836E-02} \\
    & $\VOdouFirstPi$     & \num{-3.12107139E-02} \\
    & $\VOdouSecondPi$    & \num{-2.19602031E-02} \\
    \hline
    \multirow{4}[0]{*}{$q$ } &    $\VOAPi$     & \num{-2.09279299E-04}    \\
    & $\VOBPi$     & \num{-1.63337099E-04} \\
    & $\VOdouFirstPi$     & \num{-2.04345034E-04} \\
    & $\VOdouSecondPi$    & \num{1.00240289E-03} \\
    \hline
 \end{tabular}
\end{table}

\subsubsection{Off-diagonal spin-orbit couplings}

We can observe obvious avoided-crossing structures from
the rovibronic energy distributions of $\VOBPi - \VOdouSigmap$ 
and $\VOCSigma - \VOdouSecondPi$ coupled states.
Thus, off-diagonal spin-orbit coupling curves,
whose parameters are given in Table\,\ref{tab:offspinorbit},
were introduced to model those structures in crossing regions.
The $\VOBPi - \VOdouSigmap$ spin-orbit coupling term was 
 optimized using \duo,
while the $\VOCSigma - \VOdouSecondPi$ coupling
was manually given an estimated value.
See the next section where the results are discussed.

\begin{table}[h]
    \centering
    \caption{Polynomial parameters for the off-diagonal spin-orbit coupling curves.}
    \label{tab:offspinorbit}
    \begin{tabular}{ll}
    \hline
    State & $a_0$ [\si{\per\cm}] \\
    \hline
    $\VOBPi - \VOdouSigmap$         & \num{21.5107379} \\
    $\VOCSigma - \VOdouSecondPi$     & 10 \\
    \hline
    \end{tabular}
\end{table}
  
\section{Results}

\subsection{Uncoupled states}

\begin{figure}
    \centering 
    \includegraphics{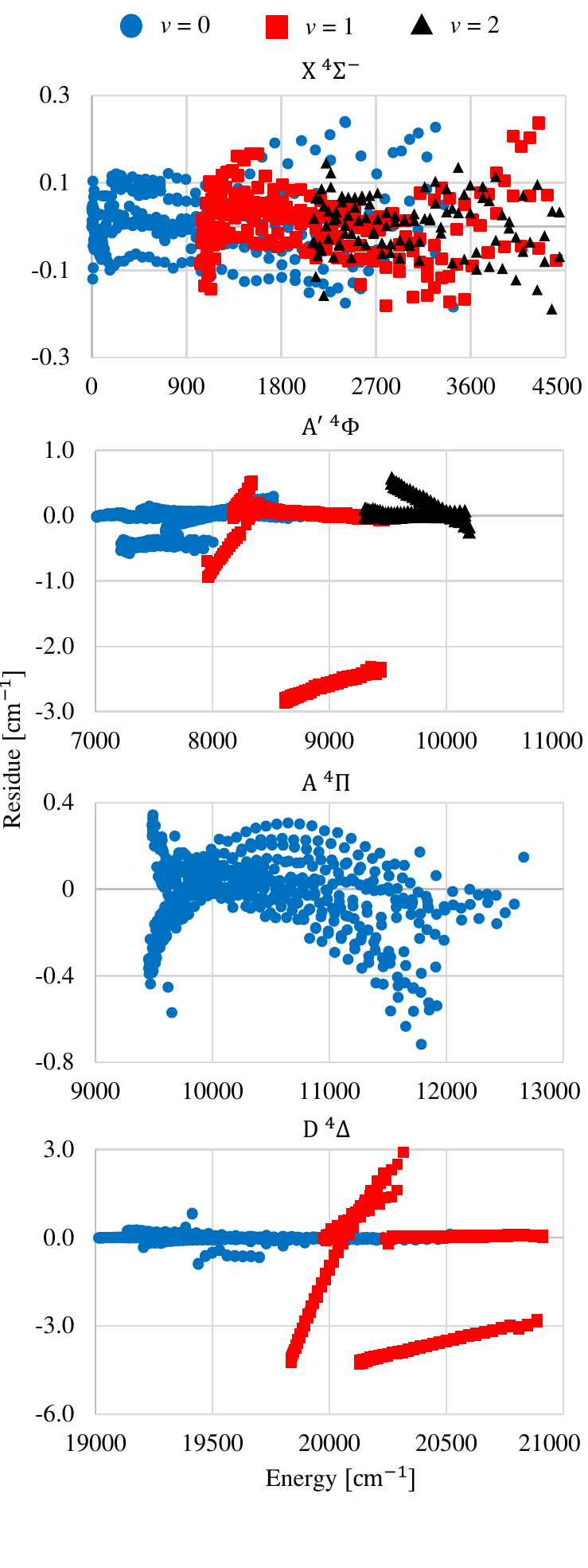}
    \caption{Fitting residues of four uncoupled quartet states:
     $\VOXSigma$, $\VOApPhi$, $\VOAPi$ and $\VODDelta$.}
    \label{fig:quarteterr}
\end{figure}

The fitting residues of the uncoupled quartet states
are given in Fig.\,\ref{fig:quarteterr}.
The energy levels of $\VOXSigma$ and $\VOAPi$
are accurately reproduced.
Compared with the $\VOXSigma$ and $\VOAPi$ states,
the spin-splittings of $\VOApPhi$ and $\VODDelta$ 
are not well estimated
which may arise from electronic state interactions not included in our model.
However,
the assignment will not be affected
as the energy gaps due to spin-orbit couplings of
$\VOApPhi$ and $\VODDelta$
are usually more than a hundred wavenumbers.

\begin{figure}
    \centering 
    \includegraphics{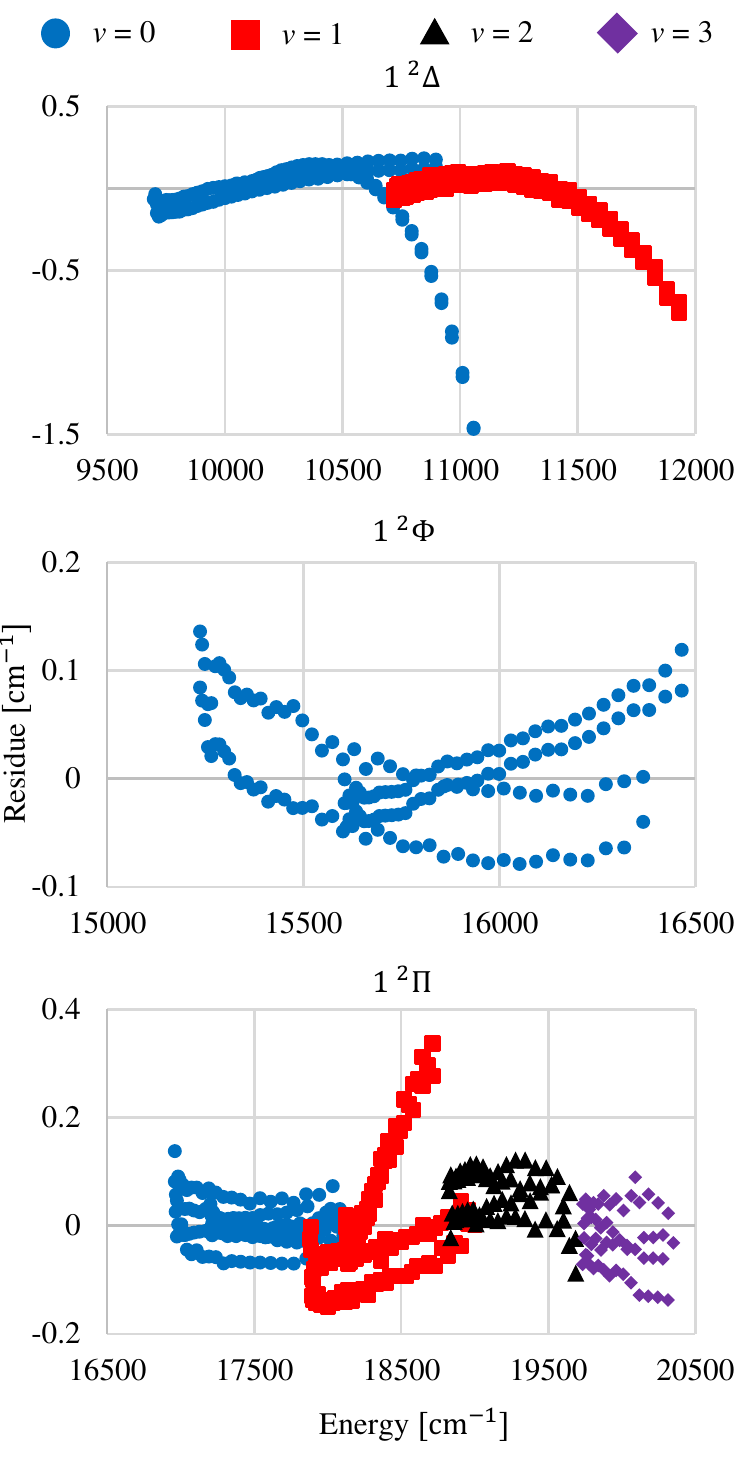}
    \caption{Fitting residues of three uncoupled doublet states:
     $\VOdouDelta$, $\VOdouPhi$ and $\VOdouFirstPi$.}
    \label{fig:doubleterr}
\end{figure}

The fitting residues of the uncoupled doublet states
are shown in Fig.\,\ref{fig:doubleterr}.
The doublet states only have two fine structure series which
makes it  easier to get good refinement results.
The errors for the high-$J$ levels of $\VOdouDelta$
increase dramatically
which is also due to state interactions.

\subsection{Coupled states}
\begin{figure*}
    \centering 
    \includegraphics{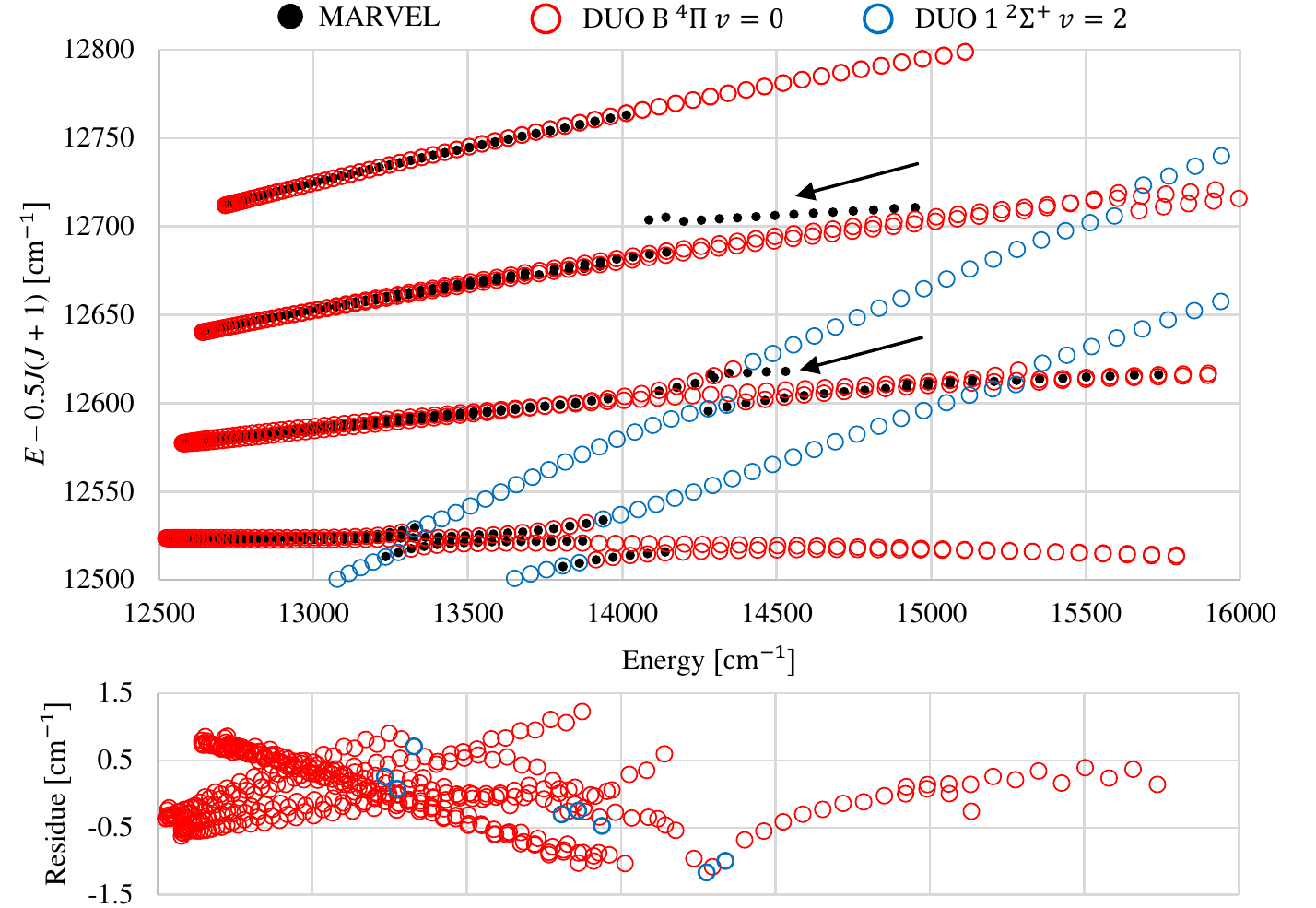}
    \caption{State interaction of $\VOBPi, v=0$ and 
    $\VOdouSigmap, v=2$. 
    The black dots in the top panel are the rovibronic energy levels 
    obtained from MARVEL analysis \cite{jt869}
    while the circles are the fitting results of \duo.
    The black arrows indicate some unreasonable data 
    which were excluded in our fitting.
    The bottom panel shows the fitting residues.}
    \label{fig:B0Sigma2}
\end{figure*}

\begin{figure*}
    \centering 
    \includegraphics{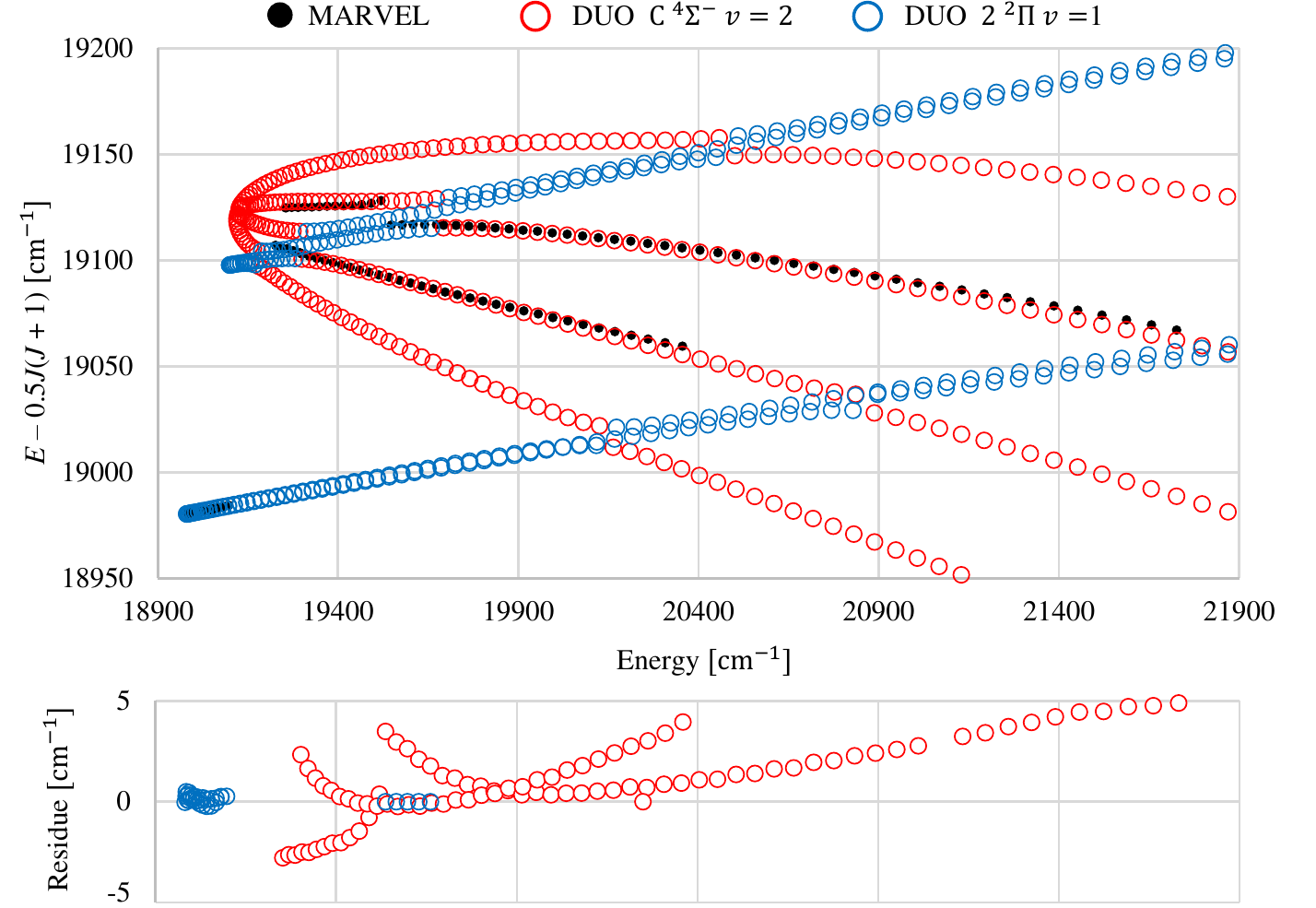}
    \caption{State interaction of $\VOCSigma, v=2$ and 
    $\VOdouSecondPi, v=1$. 
    The black dots in the top panel are the rovibronic energy levels 
    obtained from MARVEL analysis \cite{jt869}
    while the circles are the fitting results of \duo.
    The bottom panel shows the fitting residues.}
    \label{fig:C2Pi1}
\end{figure*}

In our model,
$\VOBPi - \VOdouSigmap$ 
and $\VOCSigma - \VOdouSecondPi$
are coupled by spin-orbit terms
to give reasonable avoided crossing structures,
as shown from Fig.\,\ref{fig:B0Sigma2} to Fig.\,\ref{fig:C2Pi1}.

\begin{figure*}
    \centering 
    \includegraphics{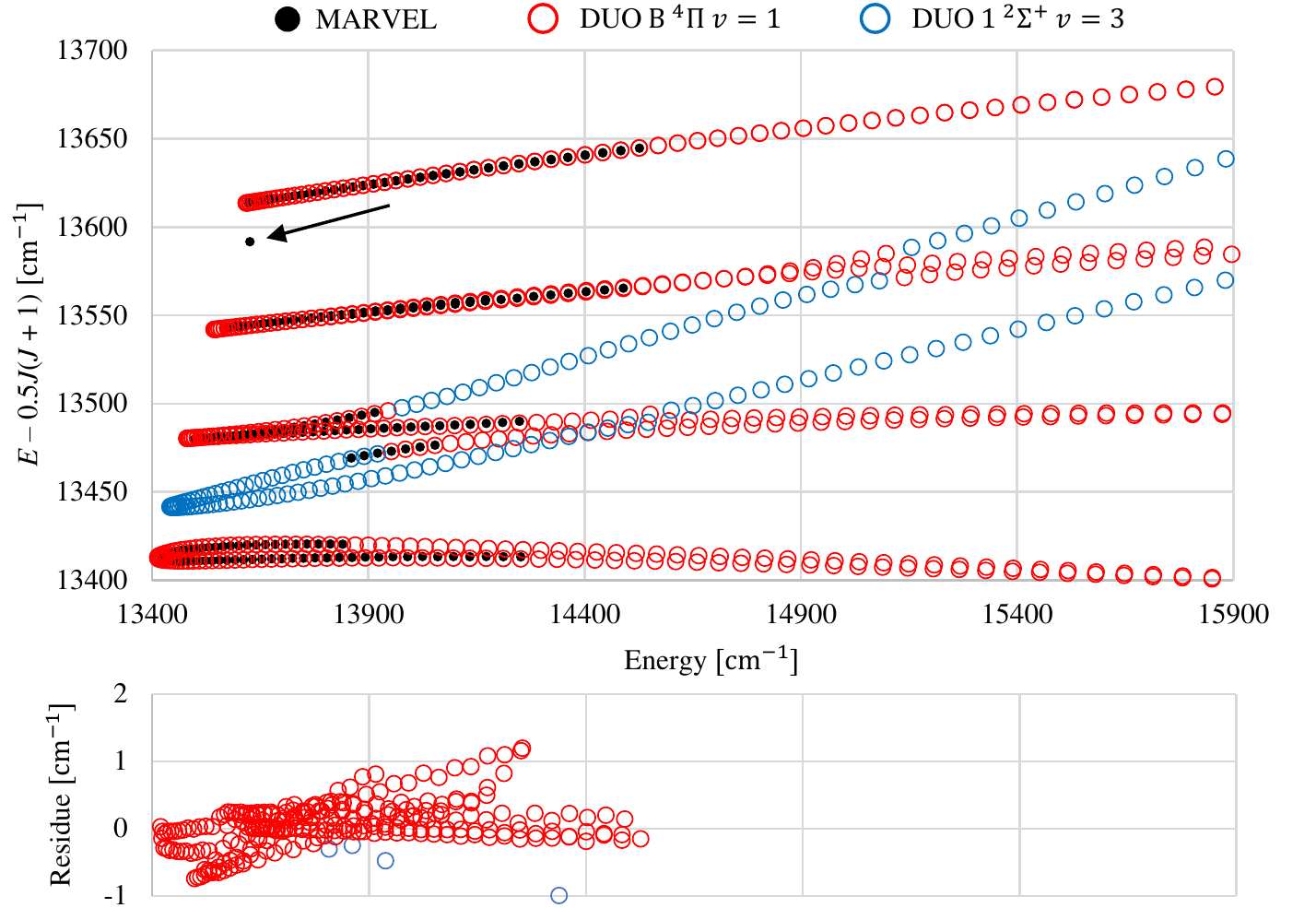}
    \caption{State interaction of $\VOBPi, v=1$ and 
    $\VOdouSigmap, v=3$. 
    The black dots in the top panel are the rovibronic energy levels 
    obtained from MARVEL analysis \cite{jt869}
    while the circles are the fitting results of \duo.
    The bottom panel shows the fitting residues.}
    \label{fig:B1Sigma3}
\end{figure*}

Figures \ref{fig:B0Sigma2} and \ref{fig:B1Sigma3}
demonstrate the smooth transitions
from on electronic state to another 
in both $e$ and $f$ series.
Note that,
the electronic state quantum numbers of 
the energy levels near the interaction region 
are  not well-determined.
They are assigned according to the dominant
contribution of the corresponding
basis function to the rovibronic wavefunction.
The MARVEL energy levels indicated by the black arrows
in these figures were excluded from the fits.
They seems belong to another electronic state 
but were not recognized in the current model.

\begin{figure*}
    \centering 
    \includegraphics{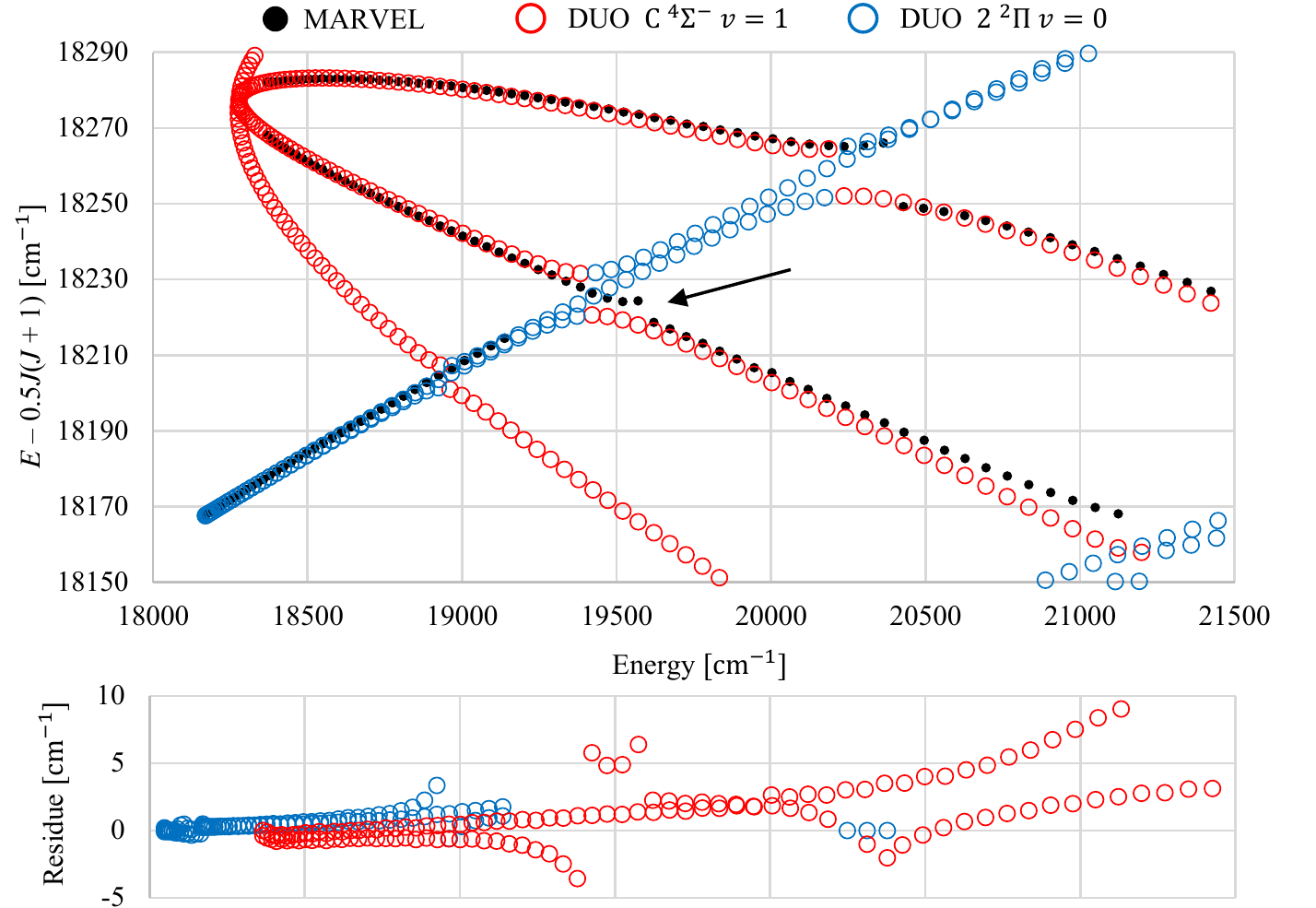}
    \caption{State interaction of $\VOCSigma, v=1$ and 
    $\VOdouSecondPi, v=0$. 
    The black dots in the top panel are the rovibronic energy levels 
    obtained from MARVEL analysis \cite{jt869}
    while the circles are the fitting results of \duo.
    The rotational constant of $\VOdouSecondPi, v=0$ in 
    derived from our model is consistent with
    that of the recently assigned results 
    (see the blue circles in the top panel).
    However, it crossing point with the $F_2$ series of $\VOCSigma, v=1$
    is inconsistent with experimental values
    (see the area pointed by the black arrow).
    The bottom panel shows the fitting residues.}
    \label{fig:C1Pi0}
\end{figure*}

The  off-diagonal $\VOCSigma - \VOdouSecondPi$ spin-orbit coupling term 
 was not fitted but
visually estimated to give an acceptable avoided-crossing 
structure as shown in Figs. \ref{fig:C1Pi0} and \ref{fig:C2Pi1}.
The $v=0, F_2$ levels of $\VOdouSecondPi$
shown in Fig.\,\ref{fig:C1Pi0} are recently
assigned in the work of Bowesman \etal \cite{jt869}.
However, the crossing point
of this series with $F_2$ series of $\VOCSigma$
is different from the previously determined one,
as indicated by the black arrow in Fig.\,\ref{fig:C1Pi0}.
All transitions connecting the $\VOCSigma$ and $\VOdouSecondPi$ states
in the MARVEL spectroscopic network are 
under review and
some of them may be reassigned.
Thus,
we did not make too much effort to improve
the accuracy here.

\begin{figure}
    \centering 
    \includegraphics{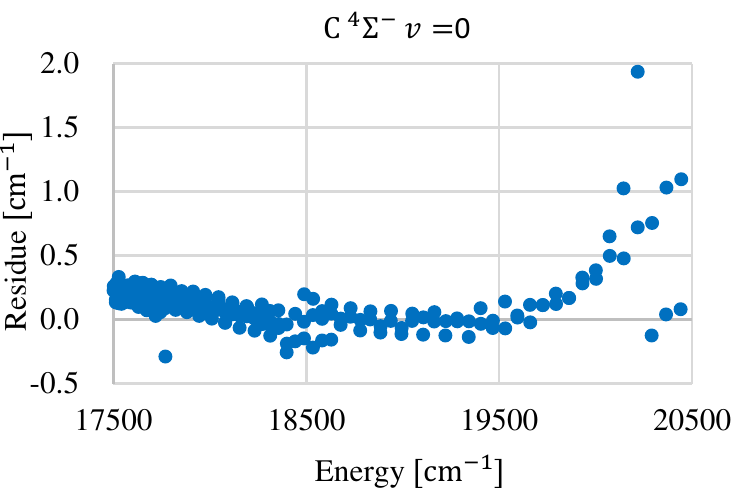}
    \caption{Fitting residues of $\VOCSigma, v=0$.}
    \label{fig:C0}
\end{figure}

The $v=0$ levels of $\VOCSigma$ do not 
interact with the $\VOdouSecondPi$ state.
Thus, 
the diagonal spin-rotation coupling term 
in Table\,\ref{tab:spinrotation}
was determined by the states with the $v=0$ levels.
Nevertheless,
there are still systematic errors as shown in Fig.\,\ref{fig:C0}.
The $J$-dependent fitting residues of high-$J$ levels 
should be attributed to $\VOCSigma - \VOdouFirstPi$
interaction, although
more experimental data are needed to verify this.

\subsection{Fitting quality}

The root-mean-square error ($\mathrm{RMSE}$)
of the fitting residues
for each electronic state is calculated as
\begin{equation}
    \mathrm{RMSE}= \sqrt{\frac{1}{N} \sum_{i=1}^N e_i}\,,
\end{equation}
where $e_i$ is the fitting error of the $i$-th energy level
and $N$ is the number of rovibronic states included.
The values of $\mathrm{RMSE}$ are listed in Table\,\ref{tab:rmse}.

\begin{table}[htbp]
    \centering
    \caption{RMSE and maximum absolute error of each state in cm$^{-1}$.}
    \label{tab:rmse}
    \begin{tabular}{cccc}
    \hline
    State & ${v} $ & {RMSE} & {Max absolute error} \\
    \hline
    \multirow{3}[0]{*}{$\VOXSigma$} & 0     & 0.0777 & 0.2404 \\
        & 1     & 0.0168 & 0.2369 \\
        & 2     & 0.0651 & 0.1884 \\
    \hline
    \multirow{3}[0]{*}{$\VOApPhi$} & 0     & 0.2241 & 0.5760 \\
        & 1     & 1.3634 & 2.8620 \\
        & 2     & 0.206 & 0.5821 \\
    \hline
    $\VOAPi$     & 0     & 0.1766 & 0.5131 \\
    \hline
    \multirow{2}[0]{*}{$\VODDelta$} & 0     & 0.1893 & 0.9044 \\
        & 1     & 2.3556 & 4.2883 \\
    \hline
    \multirow{2}[0]{*}{$\VOdouDelta$} & 0     & 0.2675 & 1.4664 \\
        & 1     & 0.2107 & 0.7573 \\
    \hline
    $\VOdouPhi$   & 0     & 0.0507 & 0.1362 \\
    \hline
    \multirow{4}[0]{*}{$\VOdouFirstPi$} & 0     & 0.0363 & 0.1380 \\
        & 1     & 0.1145 & 0.3368 \\
        & 2     & 0.0696 & 0.1217 \\
        & 3     & 0.0603 & 0.1372 \\
    \hline
    $\VOBPi - \VOdouSigmap$ &       & 0.4234 & 1.2228 \\
    \hline
    $\VOCSigma$     & 0     & 0.2949 & 1.9393 \\
    \hline
    $\VOCSigma - \VOdouSecondPi$  &       & 1.4829 & 9.0335 \\
    \hline
    \end{tabular}
  \end{table}

\begin{figure}
    \centering 
    \includegraphics{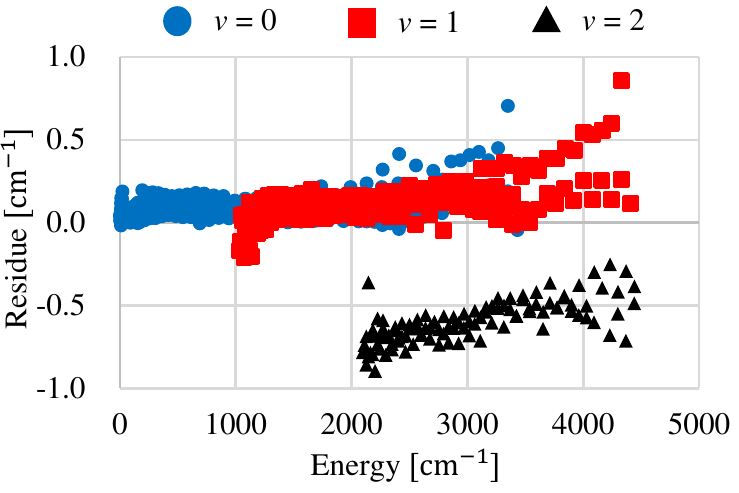}
    \caption{Energy errors for the $\VOXSigma$ state as given by
     the VOMYT model.}
    \label{fig:VOMYTXerr}
\end{figure}

The VOMYT model was refined using Model Hamiltonian, 
combination difference energy levels and
transition frequencies \cite{jt644}.
As most of the transitions connect to the ground electronic state,
VOMYT has a reasonable representation of $\VOXSigma$.
However, its vibrational structure was still
not accurately determined as shown in Fig.\,\ref{fig:VOMYTXerr}.
There is an obvious offset of the $v=2$ energy levels, 
which is likely to be significantly
worse for higher vibrationally excited states.
The $\mathrm{RMSE}$ value for the VOMYT model is \SI{0.3015}{\per\cm}, about
four times worse than ours.

The rotational constants of all states in our model 
are around \SI{0.5}{\per\cm}
and the rotational energy gaps are estimated as 
\begin{equation}
    \Delta E = 0.5(J+2)(J+1)-0.5(J+1)J=J+1 \quad \si{\per\cm}.
\end{equation}
Thus,
our model achieves sufficient accuracy to make rotational assignments.
The coupling curves can be further refined
once  revised or extended data is available.

\section{Conclusions}

We propose an empirical spectroscopic model for 6 quartet and
5 doublet states of VO.
The most recent MARVEL analysis of VO \cite{jt869}
generates accurate energy term values for this molecule,
which makes it possible to achieve experimental accuracy
for some vibronic levels
compared with VOMYT \cite{jt644}.
The interactions of
$\VOBPi - \VOdouSigmap$ and $\VOCSigma - \VOdouFirstPi$
are analyzed by introducing the off-diagonal spin-orbit
couplings between them.

While our model represents a significant advance on what is currently available, 
there remains two outstanding issues. 
The first is the treatment of resonance interactions which is discussed extensively above. 
Given that state-of-the-art \abinitio calculations on systems with a transition metal atom
are not capable of even approaching the accuracy required for high resolution studies, 
improving our model will require further experimental input.

The second issue is the hyperfine structure. 
The main isotope of vanadium, 
$^{51}$V,  has a nuclear spin $I = 7/2$ which is known
to lead to pronounced hyperfine effects in the spectrum of VO
\cite{82ChHaMe,95AdBaBeBo,08FlZi}. 
We have started to address this
issue by extending our nuclear motion program \duo\ to explicitly include
hyperfine effects \cite{jt855} within its variational model. We have
also built a hyperfine-resolved spectroscopic model for the ground $\VOXSigma$ state of $^{51}$V$^{16}$O \cite{jt873}. 
We are working on extending the model to other electronic
states of VO to give a spectroscopically-accurate, hyperfine-resolved
model of the molecule. 
Given the limited experimental 
hyperfine-resolved data available for VO, 
particularly for spectra involving excited vibrational states, 
further experimental input into this problem would be particularly welcome.

\section*{Acknowledgements}
We dedicate this article to the memory of Colin Western and acknowledge
both the help his work gave to our studies and the readiness with
which he responded to our many questions.
Qianwei Qu acknowledges the financial support from
University College London and China Scholarship Council.
This work was supported by the STFC Projects 
No. ST/M001334/1 and ST/R000476/1, 
and ERC Advanced Investigator Project 883830.

\section*{Data Availability}

The open access  programs \exocross\ and \duo\ are  available from 
\href{https://github.com/exomol}{github.com/exomol}. 
The \duo\ input file used in this work 
is given as supplementary material;
our potential energy and coupling curves 
are included as part of this input file.

\bibliographystyle{elsarticle-num} 
\bibliography{journals_phys,jtj,VO,programs,VO_MARVEL,methods, bib_exoplanets}

\end{document}